# Distortion and preservation of Giant resonances in Endohedral Atoms A@$C_{60}$

M. Ya. Amusia[1,2], A. S. Baltenkov[3], L. V. Chernysheva[2]

[1]Racah Institute of Physics, the Hebrew University, 91904 Jerusalem, Israel
[2]Ioffe Physical-Technical Institute, 194021 St. Petersburg, Russia
[3]Arifov's Institute of Electronics, Akademgorodok, 700125 Tashkent, Uzbekistan

**Abstract**

It is demonstrated in this Letter that the effect of the fullerene shell upon atomic Giant resonance decisively depends upon energy of photoelectrons, by which the resonance decay. According to the prediction in [1], the Giant resonance in Xe is strongly modified in the endohedral Xe@$C_{60}$ being transformed from a single broad and powerful maximum in Xe into four quite narrow but with almost the same total oscillator strength. On the contrary, the 4d Giant resonances in ions $Ce^{3+}$ (the electronic structure that Ce has, when stuffed into fullerene), in $Ce^{4+}$, and Eu are considered. In none of them the 4d Giant resonance in endohedrals is affected essentially. This is because the decay of the Giant resonances in these endohedrals proceeds by emission of fast photoelectrons that are almost unaffected by the $C_{60}$ shell.

The results obtained give at least qualitative explanation to the fact that recent observation of 4d Giant resonance in Ce@$C_{82}^+$, where the Giant resonance was observed as a maximum without noticeable structure.

PACS 31.25.-v, 32.80.-t, 32.80.Fb.

**1.** It has been recently predicted that the 4d atomic Giant photoabsorption resonance in Xe is strongly distorted when Xe is caged inside $C_{60}$ fullerene forming endohedral Xe@$C_{60}$ [1, 2]. Being unable due to serious technical difficulties to perform measurements on strictly this object the experimentalists measured the cross-section of the 4d photoionization of Ce@$C_{82}^+$ [3]. The results for Ce@$C_{82}$ with less detailed picture of the Giant resonance were obtained in [4], again demonstrating a maximum without noticeable structure in the region of the atomic Giant resonance.

The fullerene $C_{82}$ is a more complicated object for theoretical studies than $C_{60}$: it is likely to be deformed and the parameters of it are less know than for $C_{60}$. So, we decided to pay the experimentalists the same coin and investigate instead of Ce@$C_{82}^+$ and Ce@$C_{82}$ objects that are similar but easier for theoretical analyses. Namely, we consider in this Letter photoionization of the following objects $Ce^{3+}$@$C_{60}$, $Ce^{4+}$@$C_{60}$, and Eu@$C_{60}$. For $Ce^{3+}$ and $Ce^{4+}$ we will calculate the photoionization cross-sections for isolated ions for the first time while data on Eu will take from [5].

According to [1], the atomic 4d Giant resonance (GR) in Xe transforms into a sequence of several maxima in the endohedral atom Xe@$C_{60}$ [1, 2]. This strong distortion comes as a result of reflection of photoelectrons by the $C_{60}$. This reflection is strong, since photoelectrons from 4d in Xe are predominantly slow, coming from 4d subshell with energies of about 3Ry. The $C_{60}$ potential acts strongly upon these electrons.

It was emphasized long ago in [6], that the mechanisms of Giant photoabsorption in Eu and Xe are quite different. The electronic structure of Eu is different from that of Xe by the presence of additional half-filled $4f^7$ and filled $5s^2$ subshells. According to the Hund rule, all seven electrons in the half-filled shell have the same spin projection, for definiteness "up" or ↑. The other seven levels ("down", or ↓) are empty. The GR in Eu is a result of decay of the $4d\uparrow-4f\uparrow$ discrete excitation into continuous spectrum of the outer subshells, predominantly $4f\downarrow-\varepsilon g, d\downarrow$. The outgoing electrons' energy in Eu 4d ionization is at least twice as big as in Xe and the angular momentum higher. Therefore the effect of the fullerene cage upon the photoabsorption cross-section in Xe@$C_{60}$ is considerably smaller than in Eu@$C_{60}$. As we will show, the situation with $Ce^{3+}$@$C_{60}$ and $Ce^{4+}$@C60 is similar to that in Eu@$C_{60}$. The reason is that on the way from I and Xe, to Ce and Eu the mechanism of GR formation and decay changes at Ce. Note that we consider inside $C_{60}$ an ion, $Ce^{3+}$ or $Ce^{4+}$, not a neutral atom in accord with arguments presented in [3].

To take into account the role of $C_{60}$, in [1, 2] we used so-called "orange-skin" potential that substitutes the real $C_{60}$ potential by an infinitely thin one [7]. This approximation is valid for electrons with energies not higher than 2-3Ry. In this energy range the detailed structure of $C_{60}$ has to be inessential. In this Letter we intend to apply the same model to at least twice as high energies. There the $C_{60}$ structure and its potential have to be important, affecting the ionization cross-section and angular distribution of photoelectrons. Of course, the very fact that with growth of photoelectron energy the effect on it of the $C_{60}$ potential is evident. But on pure qualitative grounds it is unclear whether the increase of this energy by a factor of two is sufficient to make the role of $C_{60}$ shell upon the GR negligible. This requires, alas, calculations that are performed in this Letter.

Our aim here is, however, not to present the details of the cross-sections but to demonstrate that the affect of $C_{60}$ upon photoelectrons is small. One has to have in mind that account of finite thickness of the $C_{60}$ potential can only decrease and broaden the oscillations predicted in the frame of the orange-skin potential model thus decreasing the relative effect of the $C_{60}$ shell. This justifies the use of a simple model instead of a complicated real potential.

Of course, it is an oversimplification to present $C_{60}$ shell by a static, even finite thickness potential. Indeed, it has internal degrees of freedom, have collective, plasmon-type oscillations [8]. However, as it was demonstrated in [9] that the role of dynamical excitations dies out very rapidly with the growth of photon frequency $\omega$, so that at the GR frequencies the role of dynamic reaction of the $C_{60}$ electron shell can be neglected.

So, in this Letter we will limit ourselves by accounting for the $C_{60}$ in the frame of bubble-potential model.

**2.** The discussion of GRs of endohedral atoms is of great general interest and value, since Giant resonances are universal features of the excitation of any finite many-fermion systems: nuclei, atoms, fullerenes, and clusters. They represent collective, coherent oscillations of many particles and manifest themselves most prominently in photon absorption cross sections. In a nucleus Giant resonances represent the excitation of coherent oscillatory motion of all protons relative to all neutrons [10], while in all other objects mentioned above they represent the coherent motion of all electrons of at least one many-electron shell (in atoms) and all collective electrons in metallic clusters and fullerenes relative to the atomic nucleus or the positive charge of a number of nuclei. Giant resonances are manifestations of plasmon-type or Langmuir excitations in a homogeneous electron gas [11] or so-called "zero" sound in a Fermi-liquid [12].

It is important to emphasize that the ratio of the resonance width $\Gamma$ that characterizes its lifetime $\tau$, $\tau \sim 1/\Gamma$, to the frequency $\Omega$ is almost the same for all the above-mentioned objects, $1/5 \leq \Gamma/\Omega \leq 1/4$. It is obvious that the absolute values of the resonance energies and cross-sections differ in these objects by orders of magnitude, particularly when we compare the values for the $4d^{10}$ subshell in atomic Xe and nuclear Pb.

Therefore, it is of special interest and quite instructive to find a system, in which relatively small variation – substitution of one "caged" atom – Xe by another ion or atom, affect the cross-section prominently.

**3.** We will use in this Letter the theoretical approaches already developed in a number of previous papers [13-15]. However, for completeness, let us repeat the main points of the consideration and present the essential formula used in calculations.

Let us start with the problem of an isolated closed shell atom.

In one-electron Hartree-Fock (HF) approximation the $nl$-subshell photoionization cross-section by light of frequency $\omega$ in the dipole approximation $\sigma_{nl}(\omega)$ is given by the expression (see e.g. [16])

$$\sigma_{nl}(\omega) = \frac{4\pi^2}{\omega c}(2l+1)[(l+1)d_{l+1}^2 + l d_{l-1}^2] \equiv \sigma_{nl,\varepsilon l+1}(\omega) + \sigma_{nl,\varepsilon l-1}(\omega), \quad (1)$$

where the matrix elements $d_{l\pm1}$ in the so-called $r$-form are given as

$$d_{l\pm1} \equiv \int_0^\infty P_{nl}(r) r P_{\varepsilon l\pm1}(r) dr. \quad (2)$$

Here $P_{nl}(r)$, $P_{\varepsilon l\pm1}(r)$ are the radial Hartree-Fock (HF) [16] one-electron wave functions of the $nl$ discrete level and $\varepsilon l\pm1$ - in continuous spectrum, respectively.

In order to take into account the Random Phase Approximation with Exchange (RPAE) [16] multi-electron correlations, one has to perform the following substitution

$$d_{l\pm1}^2 \to \operatorname{Re} D_{l\pm1}^2 + \operatorname{Im} D_{l\pm1}^2 \equiv \tilde{D}_{l\pm1}^2 \quad (3)$$

where $\tilde{D}_{l\pm1}(\omega)$ is the absolute values of the amplitudes for respective transitions with angular moments $l\pm1$.

The following are the RPAE equation for the dipole matrix elements [16]

$$\langle v_2|D(\omega)|v_1\rangle = \langle v_2|d|v_1\rangle + \sum_{v_3,v_4}\frac{\langle v_3|D(\omega)|v_4\rangle(n_{v_4}-n_{v_3})\langle v_4 v_2|U|v_3 v_1\rangle}{\varepsilon_{v_4}-\varepsilon_{v_3}+\omega+i\eta(1-2n_{v_3})}, \quad (4)$$

where

$$\langle v_1 v_2|\hat{U}|v_1' v_2'\rangle \equiv \langle v_1 v_2|\hat{V}|v_1' v_2'\rangle - \langle v_1 v_2|\hat{V}|v_2' v_1'\rangle. \quad (5)$$

Here $\hat{V} \equiv 1/|\vec{r}-\vec{r}'|$ and $v_i$ is the total set of quantum numbers that characterize a HF one-electron state on discrete (continuum) levels. That includes the principal quantum number (energy), angular momentum, its projection and the projection of the electron spin. The function $n_{v_i}$ (the so-called step-function) is equal to 1 for occupied and 0 for vacant states.

The dipole matrix elements $D_{l\pm1}$ are obtained by solving the radial part of the RPAE equation (4) numerically, using the procedures discussed at length in [16].

In this Letter we consider open shell ion $Ce^{3+}$, closed shell ion $Ce^{4+}$ and half-filled shell atom Eu. Half-filled shell endohedral was considered in [17]. The generalization is straightforward. As it was discussed in a number of places (see, e.g. [15]), we treat such an atom as having two types of electrons, namely "up" and "down", denoted with an arrow $\uparrow$ and $\downarrow$, respectively. Due to presence of the semi–filled level $4f^7\uparrow$, each subshell splits into two levels, up" $\uparrow$ and "down" $\downarrow$ with different ionization potentials and without exchange between these electrons. As a result, each cross-section, angular anisotropy parameter, dipole matrix elements, one-electron wave functions and scattering phase of a given level become spin-dependent values $\sigma_{nl\uparrow,\downarrow}(\omega)$, $\beta_{nl\uparrow,\downarrow}(\omega)$, $D_{l\pm1\uparrow,\downarrow}(\omega)$, $d_{l\pm1\uparrow,\downarrow}$, $\delta_{l\pm1\uparrow,\downarrow}$, and $\Delta_{l\pm1\uparrow,\downarrow}$. While the photon interaction cannot connect the spin "up" and "down" states, the Coulomb interelectron interaction connects them. As a result, instead of integral linear equation (4), one has to solve a matrix integral equation that is symbolically presented in the following form:

$$\left(\hat{D}_\uparrow(\omega)\hat{D}_\downarrow(\omega)\right) = \left(\hat{d}_\uparrow(\omega)\hat{d}_\downarrow(\omega)\right) + \left(\hat{D}_\uparrow(\omega)\hat{D}_\downarrow(\omega)\right) \times \begin{pmatrix} \hat{\chi}_{\uparrow\uparrow}(\omega) & 0 \\ 0 & \hat{\chi}_{\downarrow\downarrow}(\omega) \end{pmatrix} \times \begin{pmatrix} \hat{U}_{\uparrow\uparrow} & \hat{V}_{\uparrow\downarrow} \\ \hat{V}_{\downarrow\uparrow} & \hat{U}_{\downarrow\downarrow} \end{pmatrix} \qquad (6)$$

In these same notations (4) is presented as

$$\hat{D}(\omega) = \hat{d} + \hat{D}(\omega) \times \hat{\chi}(\omega) \times \hat{U} \qquad (7)$$

**4.** Let us consider the effects of photoelectron reflection. Near the photoionization threshold they can be described within the framework of the "orange" skin potential model (see [7, 18] and references therein):

$$V(r) = -V_0 \delta(r-R). \qquad (8)$$

The parameter $V_0$ is determined by the requirement that the binding energy of the extra electron in the negative ion $C_{60}^-$ is equal to its observable value. Addition of the potential (8) to the atomic HF potential leads to a factor $F_l(k)$ in the photoionization amplitudes, which depends only upon the photoelectron's momentum $k$ and orbital quantum number $l$ [7, 18]:

$$F_l(k) = \cos\breve{\Delta}_l(k)\left[1 - \tan\breve{\Delta}_l(k)\frac{v_{kl}(R)}{u_{kl}(R)}\right], \qquad (9)$$

where $\breve{\Delta}_l(k)$ are the additional phase shifts due to the fullerene shell potential (8). They are expressed by the following formula:

$$\tan \breve{\Delta}_l(k) = \frac{u_{kl}^2(R)}{u_{kl}(R)v_{kl}(R) + k/2V_0}. \tag{10}$$

In these formulas $u_{kl}(r)$ and $v_{kl}(r)$ are the regular and irregular solutions of the atomic HF equations for a photoelectron with momentum $k = \sqrt{2\varepsilon}$, where $\varepsilon$ is the photoelectron energy connected with the photon energy $\omega$ by the relation $\varepsilon = \omega - I_A$ with $I$ being the atom A ionization potential.

Using Eq. (9), one can obtain the following relation for $D^{AC(r)}$ and $Q^{AC(r)}$ amplitudes for endohedral atom A@C$_{60}$ with account of photoelectron's reflection and refraction by the C$_{60}$ static potential (8), expressed via the respective values for isolated atom that correspond to $nl \to \varepsilon l'$ transitions:

$$D_{nl,kl'}^{AC(r)}(\omega) = F_{l'}(k) D_{nl,kl'}(\omega). \tag{11}$$

For the cross-sections one has

$$\sigma_{nl,kl'}^{AC(r)}(\omega) = [F_{l'}(k)]^2 \sigma_{nl,kl'}(\omega), \tag{12}$$

where $\sigma_{nl,kl'}(\omega)$ is the contribution of the $nl \to \varepsilon l'$ transition to the photoionization cross-section of atomic subshell $nl$, $\sigma_{nl}(\omega)$ [see (1)].

The role of polarization of the C$_{60}$ shell under the action of the photon beam at considered frequencies of about 10-12 Ry is negligible [9].

**5.** Fig. 1 presents the results of 4d Xe@C$_{60}$ photoionization cross-section. Fig. 2 gives the single-electron photoionization of Ce@C$_{82}^+$ in the area of Ce 4d [3] and it has no prominent structure.

It is written in [3]: "For Xe@C60 it was hypothesized that there are oscillations on the 4d giant dipole resonance due to reflection of the photoelectron from the fullerene cage. … the present experiment appears to exclude such a phenomenon in the metallofullerene studied. The reason for the missing structures in the $4d \to 4f$ excitation of an encapsulated cerium atom (Z=58) in comparison to xenon (Z=54) may be that the 4f subshell has already collapsed in the heavier atom and, hence, does no longer give rise to a giant resonance phenomenon".

Our view is different – the GR does exist in the photoionization cross-section but is of GR nature, namely decays by emitting fast outer shell electrons. We have estimated the probability of such decay via fullerene shell electron emission and found it to be small.

We have calculated the photoionization cross-section of 5p, 5s, 4d subshells and the extra *4f* electron in Ce$^{3+}$, the same (without 4f electron) in Ce$^{4+}$, 5p, 5s, 4d "up" and "down" electrons and 4f "up" electrons in Eu and respective endohedrals in the vicinity of the 4d threshold.

Figures 3-5 depict our results for sums of contributions of all subshells for ions Ce$^{3+}$, Ce$^{4+}$ and atom Eu and respective endohedrals in the 4d threshold region. Naturally, the C$_{60}$ parameters in the present calculations were chosen the same as in the previous papers, e.g. in [18]: $R = 6.639$ and $V_0 = 0.443$. There is no considerable influence of the fullerenes shell in the presented frequency region, as one could expect on the ground of qualitative arguments presented at the beginning of this Letter.

It is essential, however, that the area below the maximum in Fig. 2 contributes, according our estimations based on the ratio of the cross-section at the 4d-maximum in Ce@$C_{82}^+$ and $C_{82}^+$, only about one to the sum rule, while the area in Figures 3-5 is much bigger, reaching in the presented photon energy regions the following values 15.2 (14.7), 10.1 (8.09), 11.7 (10.3) for encapsulated (free) ions $Ce^{3+}$, $Ce^{4+}$ and atom Eu, respectively. We see that in all cases the calculated oscillator strength although big is about one half of the total number of electrons in ionized subshells in the caged object, including the 4d subshell.

In principle, the profound difference between measured and calculated oscillator strength can signal that emission of a single electron measured in [3, 4] is not the main channel of the GR decay. Indeed, the photoionization can be accompanied by a number of secondary processes. For instance, on the way out the photoelectron can collide inelastically with $C_{60}$ electrons thus leading to creation of double instead of single ionized endohedrals. The importance of this effect was noticed for isolated atoms long ago [19]. Example of another option is a shake-off of a $C_{60}$ electron that is a result of 4d vacancy creation. In principle multiple ionization of an endohedral can be also a contribution of an additional Auger-decay channel via the emission of $C_{60}$ electrons. This option, just as that of shake we did not estimated. We have estimated, however, also the contribution of $C_{60}$ electrons to the GR width and found it negligible.

As to our RPAE calculation, they include effectively all secondary processes that proceed on the way of the photoelectron out (see, e.g.[20]) and therefore respective cross-section obeys the sum rule.

It is evident that to bring observed and measured data into agreement further efforts from both sides – theorists and experimentalists are desired. From our point of view still the investigation of 4d resonance in Xe@$C_{60}$ in order to confirm (or to reject) the predictions in [1] is of great importance.

**Acknowledgements**

This work was supported the Hebrew University Intramural Fund.

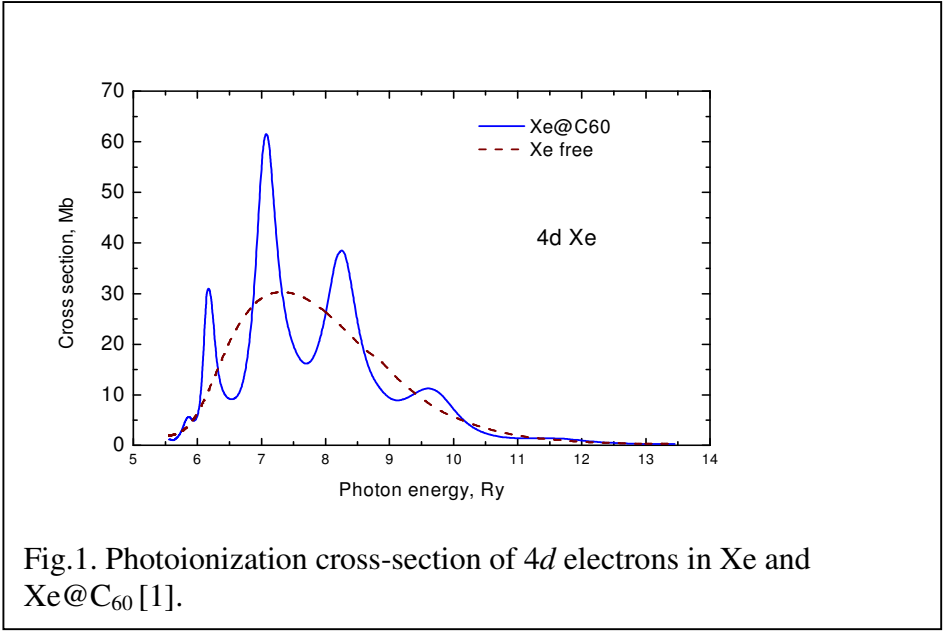

Fig.1. Photoionization cross-section of 4*d* electrons in Xe and Xe@$C_{60}$ [1].

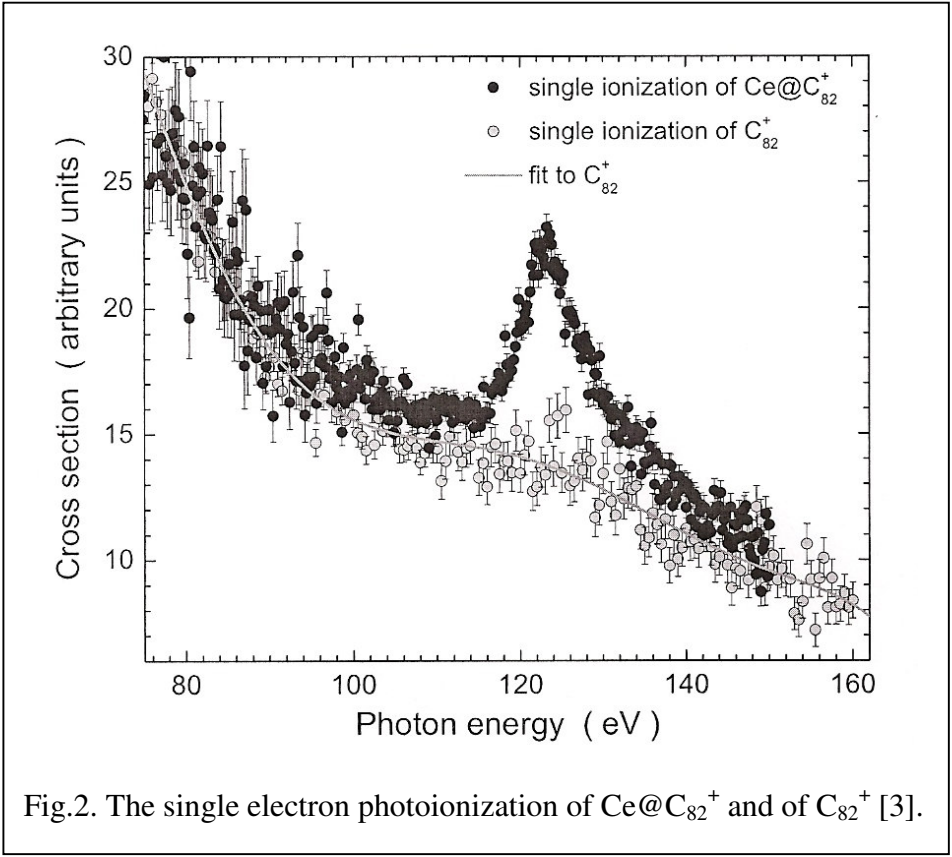

Fig.2. The single electron photoionization of Ce@$C_{82}^+$ and of $C_{82}^+$ [3].

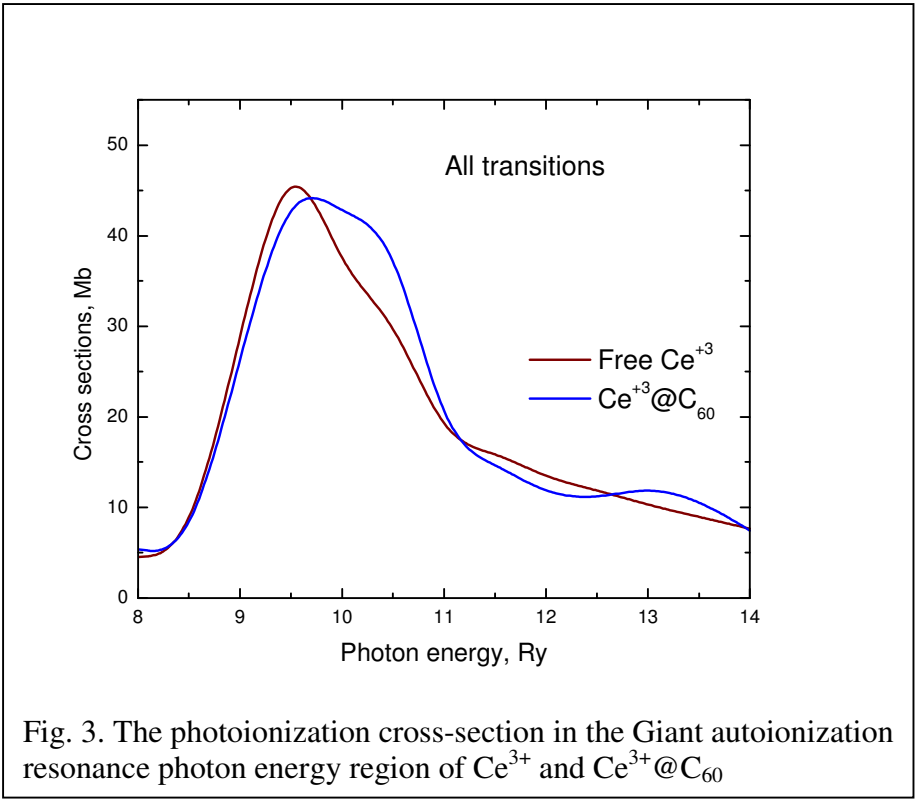

Fig. 3. The photoionization cross-section in the Giant autoionization resonance photon energy region of $Ce^{3+}$ and $Ce^{3+}@C_{60}$

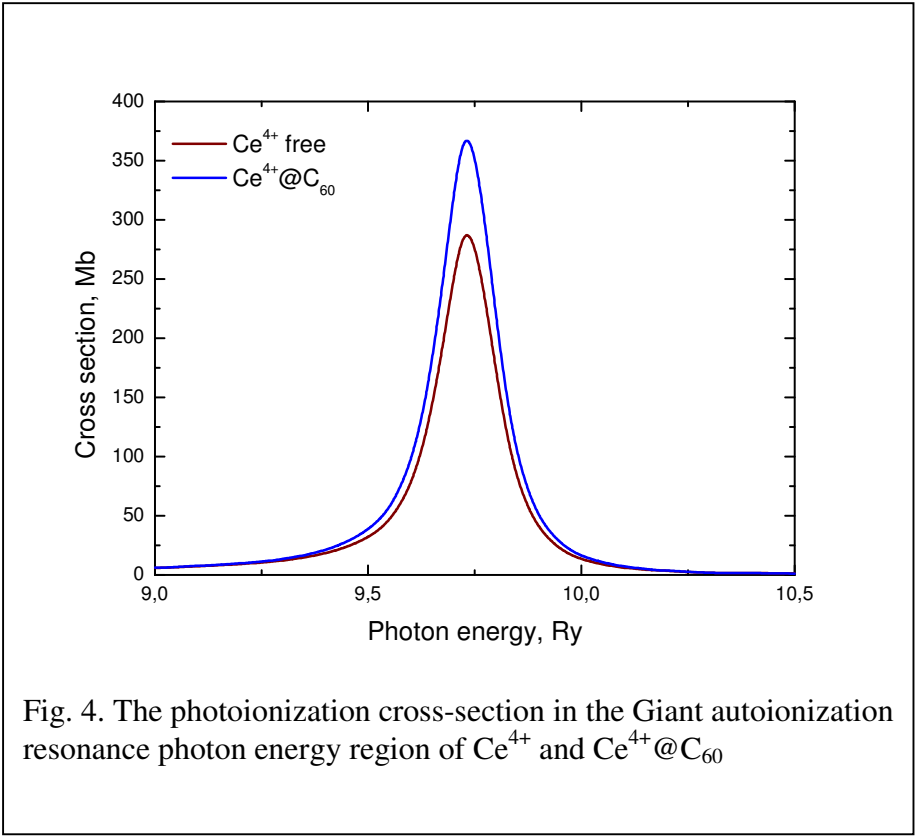

Fig. 4. The photoionization cross-section in the Giant autoionization resonance photon energy region of $Ce^{4+}$ and $Ce^{4+}@C_{60}$

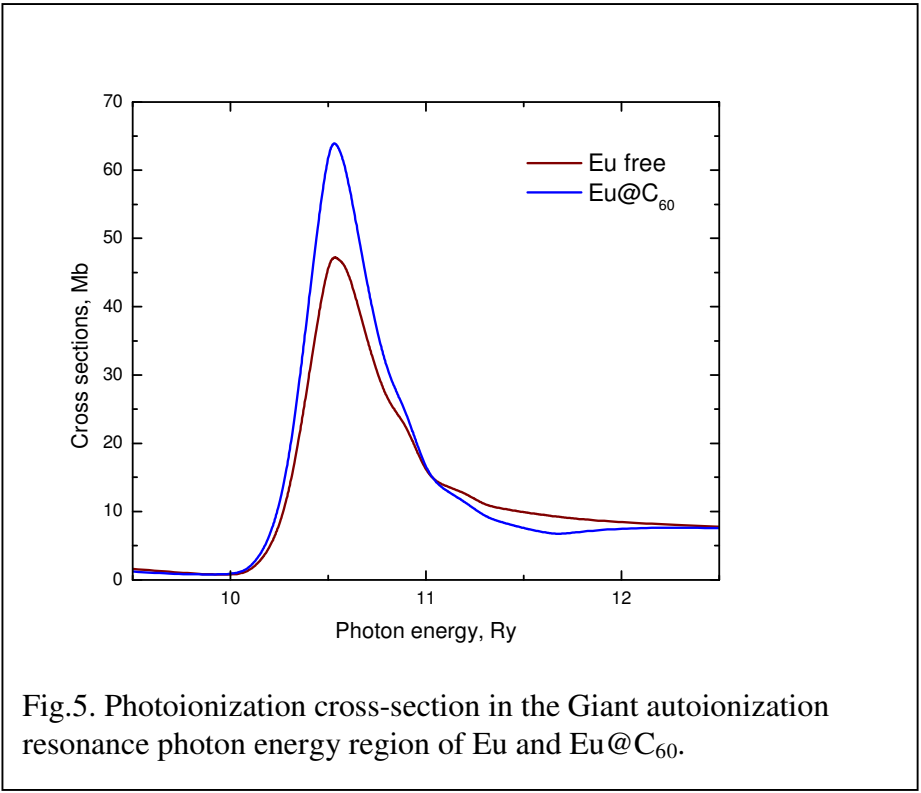

Fig.5. Photoionization cross-section in the Giant autoionization resonance photon energy region of Eu and Eu@$C_{60}$.